\documentclass[usenatbib]{mn2e}
\newcommand{\lta}{\mbox{\small\raisebox{-0.6ex}{$\,\stackrel
{\raisebox{-.2ex}{$\textstyle <$}}{\sim}\,$}}}
\newcommand{\gta}{\mbox{\small\raisebox{-0.6ex}{$\,\stackrel
{\raisebox{-.2ex}{$\textstyle >$}}{\sim}\,$}}}

\title{A multiband study of Hercules A. I. {\em ROSAT} observations of the 
intracluster medium}
\author[Nectaria A. B.
Gizani and J. P. Leahy]{Nectaria A. B. Gizani$^{1,2}$ and
J. P. Leahy,$^2$ \\ 
$^1$ Institute of Astronomy and Astrophysics, National Observatory of
Athens, I. Metaxa \& B. Pavlou, Lofos Koufou, Palaia Penteli,\\
$^{~}$ 15236 Athens, Greece.\\ 
$^2$ University of Manchester,
Jodrell Bank Observatory, Macclesfield, Cheshire SK11 9DL, UK.\\}
          
\begin{document}
\bibliographystyle{mn2e}

\maketitle

\begin{abstract}

We have made {\em ROSAT} PSPC and HRI X-ray observations to study the
intracluster gas surrounding the powerful radio source Hercules A.
The cluster is luminous in X-rays ($L_{\rm bol} = 4.8\times 10^{37}$ W),
although apparently poor in optical galaxies, and the host of the radio
source is the central dominating galaxy of the cluster.

The azimuthally-averaged X-ray surface brightness profile is well fitted 
by a modified King ($\beta$) model, 
with core radius $r_c = 121 \pm 10$ kpc and 
$\beta = 0.74 \pm 0.03$, but the cluster is elongated parallel to the radio 
source, especially on the scale of the radio lobes, and fits to 
individual quadrants give a core radius 50 per cent larger along the radio
axis. Part of this elongation appears 
to be associated with enhanced X-ray emission superimposed on the outer radio
lobes, which extend to just over $2r_c$.  There
are no obvious depressions in the X-ray emission coincident with the radio 
lobes, as expected if the relativistic plasma displaces the ICM. However,
we show that these depressions may be quite weak, essentially because
the main part of the lobes are outside the cluster core. From the surface
brightness profile for the PSPC data the X-ray emission extends out to
$\sim$ 2.2 Mpc radius. 

In the absence of the powerful jets (which must be a transient phenomenon
on cosmological timescales), we would expect a cooling flow at the centre 
of the cluster; but currently it must be substantially disturbed by the 
expansion of the radio lobes.  The PSPC spectrum reveals a cool
component of the ICM with $0.5 \lta kT \lta 1$ keV in addition to the
$\approx 4$ keV component detected by {\em ASCA} and {\em BeppoSAX}.
The central cooling time could be as low as 2 Gyr if the cool component is
centrally concentrated, otherwise it is around 6 Gyr. 
Cooling is significant on a Hubble time to a radius of about 90 kpc. 
The modelled central electron density of  $n_0 = 1.0 \times 10^{4}$ m$^{-3}$
is typical for modest cooling flows.

Finally, we have detected faint X-ray emission from a compact
central source, with size $< 15$ kpc and luminosity 
$\approx 2 \times 10^{36}$~W. 
\end{abstract}

\begin{keywords}
galaxies: active - clusters: individual: Hercules A; X-rays: galaxies; 
\end{keywords}
 
\section{Introduction}

Hercules A (3C\,348)
is the fourth brightest DRAGN\footnote{%
Double Radiosource Associated with Galactic Nucleus; 
see \citet{Leahy1993} or Leahy, Bridle \& Strom
({\tt http://www.jb.man.ac.uk/atlas/}) for a full definition.} 
in the sky at low frequencies. At a low redshift of $z=0.154$, 
its power at 178 MHz is
$P_{\rm 178\,MHz} = 1.9\times10^{27}$ W Hz$^{-1}$ sr$^{-1}$, with
$H_0 = 65$ kms$^{-1}$ Mpc$^{-1}$ and $q_0 = 0$ (used throughout
this paper).

Hercules A is identified with a very elongated cD galaxy 
\citep[e.g.][]{Sadun.etal1993,Baum.etal1996},
with absolute magnitude $-23.75$ in
the R-band \citep{Owen.etal1989}. It appears to lie at the center of a
poor, faint cluster, although measurements of the cluster richness may
be uncertain \citep*{Owen.etal1989,Yates.etal1989,Barthel.etal1996}.

Radio galaxies with the radio and optical luminosity of Her A nearly all
show hotspot-dominated Fanaroff-Riley II structure
\citep{LO96}, 
but Her A is is an exception \citep{Dreher.etal1984}.
Its structure is dominated by its twin jets, which are quite
different in appearance. The eastern one is the brightest 
(highest flux density) radio jet in the sky, and contributes a substantial
fraction of the radio luminosity of Her A; a weaker jet to the west
leads to a striking series of shells which dominate the western lobe.
There are no compact hotspots.  The structure is formally FR class I,
but does not resemble typical FR\,I objects in detail. 

X-ray emission from the Hercules A cluster was first detected by the
{\em Einstein Observatory} \citep{Feigelson.etal1983,Dreher.etal1984}.
They estimated the 0.2--4 keV luminosity as 2 $\times 10^{37}$~W. 
This is
typical of a richness 0 to 1 Abell cluster \citep{Abell1958}. In their
deprojection analysis of the {\em Einstein} data on clusters of galaxies,
\citet*{White.etal1997} list Hercules A as a possible large cooling
flow, although their best-fit model had a zero inflow rate due to
their low spatial resolution.

\citet{Barthel.etal1996} have shown that Hercules A and other radio
galaxies embedded in dense cluster gas are anomalously radio-loud 
(for their far-infrared luminosity), as expected because their confinement
is more effective, reducing adiabatic expansion.

In this first paper of a series of three,
we report observations of Hercules A in X-rays made with the {\em ROSAT}
PSPC and HRI detectors. \citet[hereafter Paper II]{Gizanic} 
present new VLA observations
of this powerful DRAGN, and we discuss the
results derived from the spectral index and projected magnetic
field/fractional polarization images all at 1$''\!$.4 resolution.
In Paper III (Gizani, Leahy \& Garrington in preparation)
we combine the results of Papers I and II to study
the magnetic field of the cluster gas surrounding the 
DRAGN. Preliminary reports have already 
been published by Gizani \& Leahy (1996, 1999).\nocite{Gizania,Gizanib}
Our observations have also been discussed, in less detail,
by \citet*{Siebert.99}.

Sections 2 and 3 describe
the observations and data reduction for the PSPC and HRI respectively.
In Section 4 we analyse the radial surface brightness profile to infer
the density distribution in the Her A cluster. In section 5 we compare
the radio and X-ray structures to search for detailed correspondences, 
and review observations of related objects to put our results into
context. Section 6 gives our conclusions.

\section{PSPC Observations and Data Reduction}
\label{pspc}

The total observing time with the PSPC detector was 8065~s, giving
a live time of 7876~s.   There were
5 observation intervals during 1993 August 19 -- 31.
These data were processed with the {\sc asterix} package, which is the
U.K. X-ray data analysis package running within the Starlink ADAM/ICL
environment.
 
The raw data, in MPE format, were pre-processed and converted to the
{\sc asterix} internal format. Then they were `cleaned' 
using standard houskeeping parameters \citep{Osborne1993}:
to ensure good aspect solutions, we only accepted times with ASP\_ERR$\leq$ 1,
which removed 3 seconds of data.
We have found no significant difference by applying screening to the
spectral analysis for low particle background using the master veto
rate (EE-MV) $< 170$. Therefore in the following analysis we have
included all data.

\subsection{Spectral Analysis}
\label{pspcbaspe}

The X-ray emission from Her A is sufficiently compact that the radial
variation in sensitivity in the {\em ROSAT} field can be neglected.
We therefore extracted a background-subtracted spectrum in {\sc
asterix} using the normal `point source' procedure. We used a source
region with radius 4.7 arcmin, 
and a background
taken from an annulus 
extending to 9
arcmin.
The background-subtracted spectrum was corrected for exposure,
dead time, vignetting, and for photons scattered out of the detect
cell (assuming a point source model).

The spectrum was processed with the {\sc ftools} program {\sc grppha}
using the binning recommended by \cite{hand}.
The data and response
function were then read into {\sc xspec} for spectral fitting.
 
We fitted a MEKAL model \citep{Kaas92} with neutral absorption using
the {\sc xspec 11.0} cross-sections \citep{Balucinska.etal1992}, 
and \citet{Anders.etal1989} relative abundances.
We do not consider $N_{H} <
5.5\times\,10^{20}$~cm$^{-2}$ as this is ruled out by H{\sc i}
absorption \citep*{Colgan.etal1988}, which detects only cold H{\sc i}
clouds and so is a lower limit. 
We use $N_{H} = 6.2\times\,10^{20}$~cm$^{-2}$ \cite{Star92} which 
is in excellent agreement with the best fit $N_{H}$ for our X-ray
spectrum. 
A single-temperature fit for $Z = 0.3$, a metallicity value
typical of clusters, which is consistent with the data, gives $kT =
2.52^{+0.52}_{-0.36}$ keV, very similar to that found by \cite{Siebert.99}. 
However,
observations at higher energies give higher values for the best-fit
temperatures: the LECS detector on {\em BeppoSAX} gives $\approx 3$
keV in the 0.2--4 keV band , while the 1.5--9.5 keV MECS detector
gives $kT = 4.8 \pm 0.6$ keV and $Z = 0.36 \pm 0.15$
\citep{Trussoni.00}. The 0.8--9 keV {\em ASCA} data of Siebert et al.
are in excellent agreement with this, but Siebert et al. prefer a fit
including a weak AGN component (consistent with the HRI image, see
below), which gives $kT = 4.25^{+1.00}_{-0.66}$ keV and Z$ = 0.44 \pm
0.13$.  At this metallicity the PSPC best-fit temperature becomes
$2.7^{+0.6}_{-0.4}$ keV, still not enough for reconciliation with {\em
ASCA} \& MECS.  Thus, as noted by Siebert et al., the combined results
suggest that there is a multiphase cluster gas around Hercules A, and
none of the single-temperature fits can be taken at face value.

\begin{figure}
\centering
\setlength{\unitlength}{1cm}
 
\begin{picture}(8.5,7)
\put(-0.5,7.5){\includegraphics{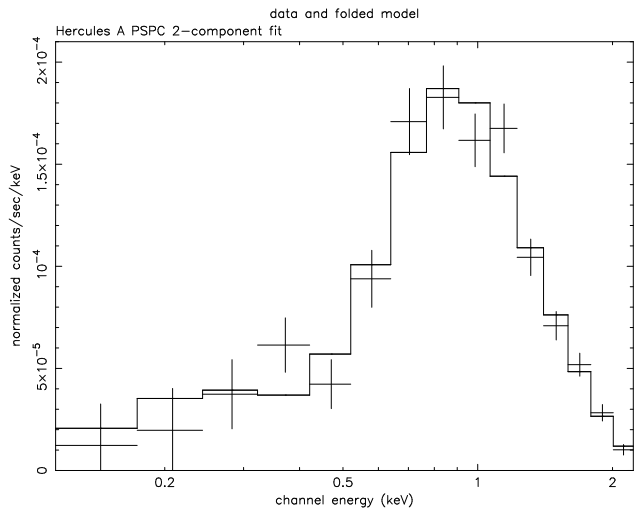}}
\end{picture}
\caption{The background-subtracted PI spectrum of Hercules~A with
the absorbed 2-temperature MEKAL model fitted. Model parameters are 
$N_H = 6.2\times 10^{20}$ cm$^{-2}$, $kT = 4.25$ and $0.69$ keV, $Z=0.44$
and $z=0.154$, relative normalization 11.8:1 (corresponding to 
hard band flux ratio 6.0:1). Reduced $\chi^{2}$=1.21, with 12 degrees
of freedom.}
\label{spec}
\end{figure}

We have therefore
tried a two-temperature fit to the PSPC spectrum with metallicity fixed
at 0.44 and the hotter component fixed at 4.25 keV. For completeness
we included a power-law component with $\Gamma = 1.9$ and normalization 
fixed to yield the point-source flux derived in our image analysis 
(see below), although this is too weak (about 5 per cent of the total 
counts) to significantly affect the fit.  
With one extra free parameter, the fit improves by 
$\Delta\chi^2 = 3.5$:
the best fit temperature for the cool component is 0.71 keV, 
with a $1\sigma$ range 0.44--1.01 keV; Fig.~\protect{\ref{spec}} 
shows this model overlaid on the observed spectrum for Hercules A.
Although such a two-component
model must be over-simplified, the PSPC response varies little with 
temperature or metallicity for gas above 3.5 keV, so
the components plausibly stand for a $\sim$4 keV hot
matrix, and a cooling-flow region, which can clearly comprise a range of
temperatures given the uncertainty in the best-fit value. The cool gas
could also be associated with the radio lobes, as seen in {\em XMM} and
{\em Chandra} observations of Vir~A \citep{Belsole01}, 
Hyd~A \citep{Nulsen02}, 
and Per~A \citep{Allen.et00}, in each of which
the radio jets seem to be dredging low-entropy gas out of the
cluster centre.
The excellent fits of single-temperature 
models to the MECS and {\em ASCA} spectra do suggest that there is 
little gas at 
intermediate temperatures. The cool component contributes 11--21 per cent to
the {\em ROSAT} count rate (excluding the central compact source) with
the higher contributions at the higher temperatures, and little dependence
on metallicity. However, because it is
dominated by line emission, its normalization in terms of emission
measure depends strongly on metallicity.

The absorption-corrected bolometric luminosity of the Hercules A
cluster is L$_{\rm bol} \simeq 4.8 \times 10^{37}$ W (see
Section~\ref{surf}). The $kT-L_{\rm bol}$ relation for cooling
flow clusters of \citet{AF98} implies a temperature of around 3.6 keV
for such a luminosity (with quite large uncertainty),
consistent with the {\em ROSAT}, {\em BeppoSAX} and {\em ASCA} results.

Unfortunately in our PSPC data there are too few counts outside 
the central resolution element to allow meaningful determination of the 
spectrum as a function of radius, so we cannot
investigate any radial temperature gradients. Radial spectral analyses
of the {\em ASCA} and {\em BeppoSAX} 
data by Siebert et al. and Trussoni et al. were
also inconclusive, unsurprisingly in view of the lower resolution of these
instruments. We show below that the cool gas could represent the majority
of the emission from the core of the cluster, but the phases could also
be mixed, perhaps as a result of the expansion of the radio lobes.

We derive the gas density from radial profile fits to our X-ray images.
Following \citet{Leahy.etal1999} we evaluated the conversion
factor between brightness and emission measure,
for our assumed `hot' component:
for $S$ in ct\,s$^{-1}$ in the PSPC hard band (PI channels 52--201):
\[ f_C(T,Z,z) = 2.05 \times 10^{-24} {\rm \; m^5\,ct\, s^{-1}\, sr^{-1}}, \]
The conversion factor between ct\,s$^{-1}$ and W\,m$^{-2}$ (0.1--2.4 keV,
absorbed)
is $f_E/f_C = 1.55 \times 10^{-14} {\rm \; W m^{-2}/\,ct\, s^{-1}}$, and
the cooling function, giving the conversion factor for 
`bolometric' (0.01--100 keV) emissivity is:
\[ \Lambda(T,Z) = 1.90 \times 10^{-36} {\rm\; W\,m^{3}}. \]

Our model gives
\begin{equation}
I / ({\rm ct\, s^{-1}\,sr^{-1}}) 
= 6.31 \times 10^{-5} EM / ({\rm m^{-6}\,kpc}),
\end{equation}
where $EM\,=\,\int n_{e}n_H\,dl$ is the emission measure.   For
metallicities in the observed range for clusters, $0.16< Z < 0.5$
(e.g. Scharf \& Mushotzky 1997), \nocite{Scharf.etal1997} $f_C(T,Z)$
varies by less than 10 per cent from our quoted value for $2.4 < kT <
10$~keV. However  $f(T,Z)$ for the cool component increases rapidly 
with metallicity;
for $Z=0.5$ it reaches twice our quoted value for $0.5 < kT < 1$ keV.
If the cool gas is concentrated in the cluster core it likely has 
higher-than-average metallicity and densities will be {\em over-}estimated
in that region by up to 50 per cent, but nearly correct at larger radii; 
if it is distributed throughout the cluster with a similar metallicity to 
the hot gas, the average density will be over-estimated by a few percent.

\subsection{The PSPC Image}
\label{hi}

Fig.~\ref{spec} shows that, due to Galactic neutral hydrogen absorption,
there are few source photons in the channels below 0.5~keV; on the
other hand the background is highest in these channels.
Therefore the signal-to-noise ratio (SNR) is much higher in
the hard band, 
and therefore we confine image analysis to this band. This has the
added benefit that the PSF is smallest in this
band and varies relatively little with energy.
We created a background subtracted hard-band image 
using the
same region as for the spectral analysis, and corrected
for exposure, dead time, and vignetting. 
The position scale
was set by aligning the X-ray peak with the radio core. The offset
of about 6 arcsec is consistent with the known pointing errors
of {\em ROSAT}.

\begin{figure}
\centering
\setlength{\unitlength}{1cm}
 
\begin{picture}(8.5,8.5)
\put(-0.6,-2.3){\includegraphics{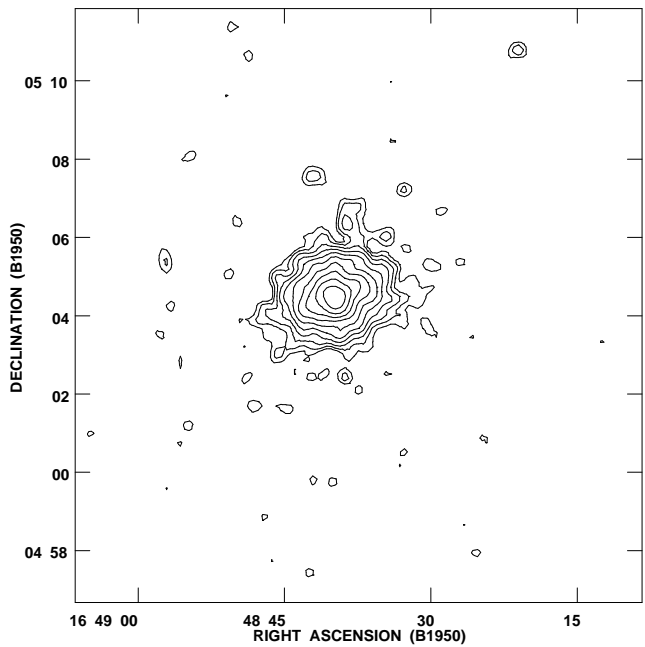}}
\end{picture}
\caption{The 
PSPC image of Hercules A. 
The image has been smoothed with a 20-arcsec FWHM gaussian. Resolution is
32 arcsec FWHM, including the instrumental PSF. The image was projected
into B1950.0 coordinates in {\sc aips}. The first contour is at
$5.29 \times 10^{-10}$ W\,m$^{-2}$\,sr$^{-1}$, and contours are separated
by ratios of $\sqrt{2}$. The background is
$\approx 1.6 \times 10^{-10}$ W\,m$^{-2}$\,sr$^{-1}$ 
$\equiv 0.9$ ct beam$^{-1}$.}
\label{xraypspc} 
\end{figure}

The resulting image, smoothed with a gaussian of FWHM 20 arcsec, is
presented in Fig.~\protect{\ref{xraypspc}}. 
Note that the fractional uncertainty in brightness in a photon image
smoothed with a gaussian is $\sqrt{1/2N}$ \citep*{HardcastleW98}, where
$N$ is the number of photons within the smoothing effective area
(hereafter `beam') $A=1.13({\rm FWHM})^2$. In Fig.~\ref{xraypspc},
$N\approx 0.9$ for the background, so the errors for faint emission
are highly non-gaussian.

\subsection{The Radial Brightness Profile}
\label{radpro}

The faint outer halo of the Hercules A cluster is best seen in the radial 
brightness profile; however in this case the background is detected with good 
signal-to-noise and it was necessary to take more care over 
background subtraction before we could obtain an accurate profile.
In order to allow the background to be fitted as part
of the overall profile fitting, our procedure was as follows.

We located the clearly-detected point sources in the 40-arcmin diameter
central portion of the PSPC field, using 
a simple peak finder applied to the image in Fig.~\ref{xraypspc}. Peaks
within 5 arcmin of the cluster centre were not included in the list as
they are likely to be substructure or random fluctuations in the cluster
gas (the only clear point source in this region in our HRI image is
not detected by the PSPC).
We created a mask consisting of discs with radii ranging from 25 to
90 arcsec (depending on distance from the field centre, to allow for
aberration) centred on each source. This was applied to the raw data,
and a radial profile  extending out
to 0\fdg 3 was extracted, by integrating in rings around the 
cluster centroid using {\sc iradial}, with a radial bin size of 2.5 arcsec.

The raw data must be corrected for vignetting, exposure and dead time. 
The correction as a function of radius is not explicitly available from
{\sc asterix}, so we derived it by taking the ratio of profiles derived
from the corrected image (from {\sc xrtcorr}) and the image from {\sc xrtsub}.
To model the vignetting of the background, 
we also made a profile of the `background'
image created by {\sc xrtsub}. (The same mask was applied in all cases).
The model background includes a small (1.3 per cent) unvignetted
contribution from the particle background, but this negligible in our case.
The model was derived from a `spectral image' of the background annulus, 
in which the band was divided into 
5 channels to accurately account for the energy-dependent radial vignetting. 
Since this region contains some source emission, we regarded the
amplitude of the background as a free parameter in our fitting procedure
(Section~\ref{surf}).
All profiles were converted to ascii files using {\sc ast2txt}, and
further processed with our own software. 

The correction
was applied to the raw data, to the associated errors (initially 
assumed equal to poisson errors), and to the background model. 
Photons from an on-axis point source will not be affected by vignetting,
so to compensate for the application of the correction to all the photons,
we also applied a correction (normalized to unity at the centre)
to our PSF model.

There is a slope in the background across the PSPC field of view, with
an amplitude of $\sim \pm 15$ per cent at 15 arcmin from the centre; 
however this is cancelled by azimuthal averaging.  
To get an idea of smaller scale
fluctuations that might affect our profile, we
summed the masked and corrected image between 0\fdg 15 and 0\fdg 3 in six
radial bins; the rms between bins was 4 per cent of
the (unsubtracted) background, consistent
with expected poisson fluctuations. 
We note that each radial bin
has a similar solid angle to the sky region occupied by the cluster.
Since it would be fortunate if the large-scale structure in the background
were purely a linear slope, we conservatively assume there may be
systematic fluctuations in the azimuthally-averaged background at the 
3 per cent level.

The profile was binned to give binwidths $\ge 5$ arcsec, and 
a signal-to-noise ratio (SNR) of $>$5:1, subject to a maximum
bin width of 90 arcsec. 
The SNR was calculated after background subtraction, and
the error used included our 3 per cent systematic uncertainty, added
in quadrature to the poisson uncertainties. All bins contained $>25$ photons
before background subtraction. 

The binning and fitting procedure (Section~\ref{surf})
was iterated a couple of times to ensure self-consistency.
The final result is shown as the starred points in Fig.~\ref{pshri}. 

\section{HRI data Analysis}
\label{hri_data}
Our HRI observation was split into two periods separated by six months.
The first, with a duration of only 1232 s, was on 1996 February 18.
The second, with the remaining 21516~s,
contained 12 observational intervals during 1996 August 27--30.
Observations taken six months apart have opposing spacecraft orientations and
frequently show large pointing offsets (D. Harris, priv. comm.). 
Since there were no strong sources
in the field to allow accurate alignment, we decided to analyse only the
second dataset. We first corrected it for the timing error in the standard 
analysis which affected the aspect solutions (using a  
script distributed by the Harvard-Smithsonian Centre for Astrophysics).
The data was processed in {\sc asterix} and {\sc ftools}.

Because of the
uncertainty in spectral response of the HRI \citep[e.g.][]{Greiner1999}, 
we did not attempt a 
spectral analysis. However, we did analyse the pulse-height analyser (PHA)
distribution of source and background to try to optimise the signal-to-noise
ratio (SNR). The background was determined in the annulus between 4 and 7
arcmin around Her A, which does not contain any background sources.
The high HRI background makes
it much less sensitive than the PSPC to very faint emission, 
so we decided to optimise the HRI image for the moderately bright emission
around the radio lobes (at radii 50--100 arcsec), giving channels 3--7 as
the best choice. These channels contain 83.3 per cent of the counts in a 
background-subtracted PHA spectrum of the source.

We made an image with the full spatial resolution (0.5-arcsec pixels)
using channels 3--7, subtracted the background (0.00464 ct pixel$^{-1}$,
derived from the profile fitting in Section~\ref{surf}),
and converted to count rate by dividing by the exposure time. No vignetting
corrections were applied but these are unimportant within 5 arcmin
(the background is dominated by charged particles which suffer no
vignetting). We also did not correct for detector quantum efficiency 
variations, as these are not well determined at high resolution; we
include a 4 per cent systematic error in the background to allow for this
(c.f. Leahy \& Gizani 2001) \nocite{Leahy.etal1999}.

We converted from observed count rate to emission measure in the same way as 
for the PSPC data, except that we used the observed PHA spectrum to estimate 
the fraction of the HRI counts that fall into our selected range of PHA 
channels. For our best fit model and our chosen PHA bands, the HRI is 
2.86 times less sensitive than the PSPC, 
and this ratio varied by less than 2 per cent for any of the plausible
models we tried.
The lower sensitivity of the HRI is cancelled by the longer exposure time, 
so that the two images have nearly the same sensitivity to bright emission.

\begin{figure*}
\centering
\setlength{\unitlength}{1cm}
 
\begin{picture}(17.5,13.2)

\put(-1.1,14.3){\includegraphics{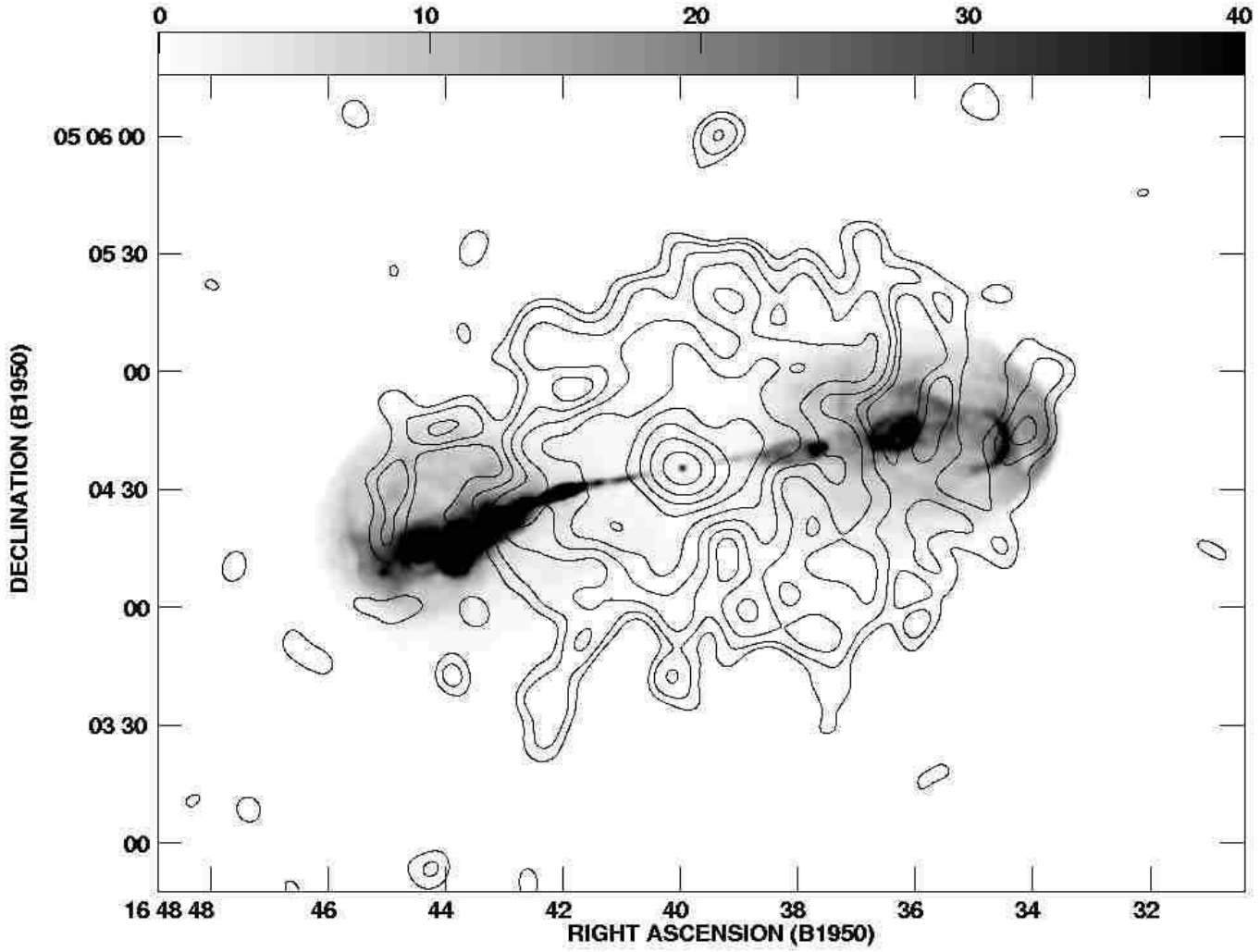}}
\end{picture}
\caption{A grey-scale image of the Hercules A radio emission at 1.5 GHz
with contours of the HRI X-ray image superimposed. The radio image has
resolution 1.4 arcsec FWHM, 
and the greyscale range is 0 to 40 mJy beam$^{-1}$.
The HRI image has been smoothed with a gaussian of FWHM 10 arcsec, 
giving an effective resolution of 11.5 arcsec including the instrumental PSF.
Contours are separated by factors of $\sqrt{2}$ starting at 
$2.37 \times 10^{-9}$ W\,m$^{-2}$\,sr$^{-1}$ $= 3.07$ ct beam$^{-1}$
($3\sigma$ above the background).
The background level (which has been subtracted)
is $1.63 \times 10^{-9}$ W\,m$^{-2}$\,sr$^{-1}$ 
$= 2.10$ ct beam$^{-1}$.}
\label{hrihigh}
\end{figure*}

Fig.~\protect{\ref{hrihigh}} shows a greyscale of our $\lambda$20-cm 
radio image 
(from Paper II) with contours from the HRI, with 10-arcsec smoothing,
superimposed. 
Positions were again set by aligning with the radio; the HRI peak (after
smoothing) was within 2 arcsec of the radio core. The HRI peak is not
completely symmetric, but we suspect that this is due to
aspect errors during the observation. 

\begin{figure}
\centering
\setlength{\unitlength}{1cm}
 
\begin{picture}(8.5,12)
\put(-0.8,-0.4){\includegraphics{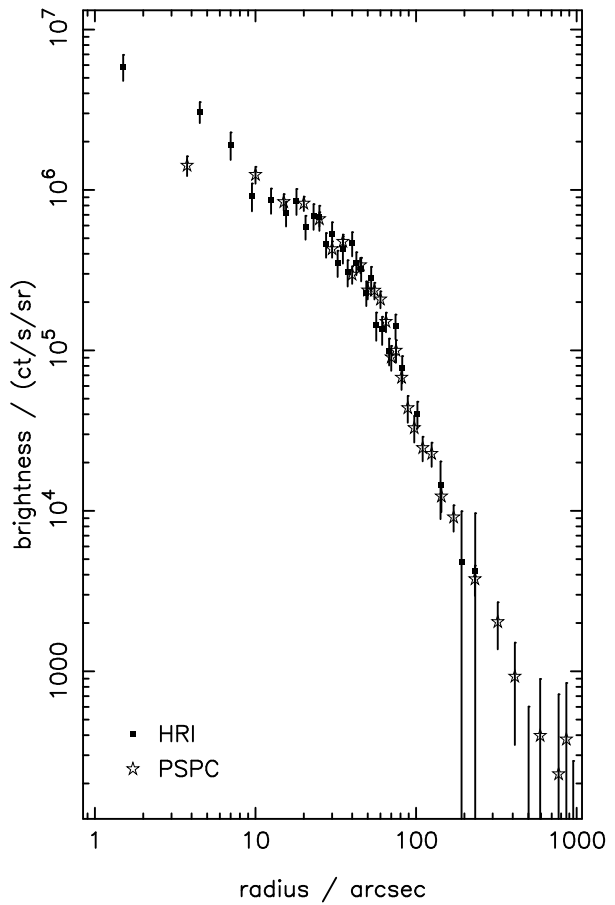}}
\end{picture} 
\caption{The X-ray surface brightness profile of the Hercules A
cluster from {\em ROSAT} PSPC (stars) and HRI (squares), after backgrounds
have been subtracted. Error bars are 1$\sigma$
allowing for 4 per cent and 3 per cent uncertainty in the HRI and
PSPC backgrounds respectively.}
\label{pshri} 
\end{figure}

The radial profile for the HRI data was created using {\sc iring} in
{\sc aips} as there were no spatially-varying corrections to be
applied.  The raw bin size was 1 arcsec and the profile extended to
250 arcsec.  Full details of the procedure are described by Leahy \&
Gizani (2001).  The profile was binned in the same way as for the
PSPC, except that we assumed a 4 per cent systematic error in the
background and used minimum and maximum bin widths of 2 and 50 arcsec
respectively.  The background-subtracted profile is plotted in
Fig.~\protect{\ref{pshri}}, multiplied by 2.86 to put it on the same
scale as the PSPC data.

There are superficial differences between the PSPC image and the HRI
image convolved to the same resolution, but these are approximately
consistent with random noise: the apparent extension of the cluster
peak emission, more or less perpendicular to the radio axis 
(see Fig.~\ref{xraypspc}) is not confirmed by the HRI data. Whereas
the PSPC seems to find a deficit of flux at $r\approx 100$ arcsec relative
to a $\beta$ model (see below), the HRI image shows a (marginally
significant) plateau at about this radius, especially to the
south-east, and in the `spur' extending to the north of the cluster,
(which is considerably more diffuse in the smoothed HRI image than in
Fig.~\ref{xraypspc}).

\begin{figure*}
\centering
\setlength{\unitlength}{1cm}
 
\begin{picture}(17.5,15)
 
\put(-1.8,15.5){\includegraphics{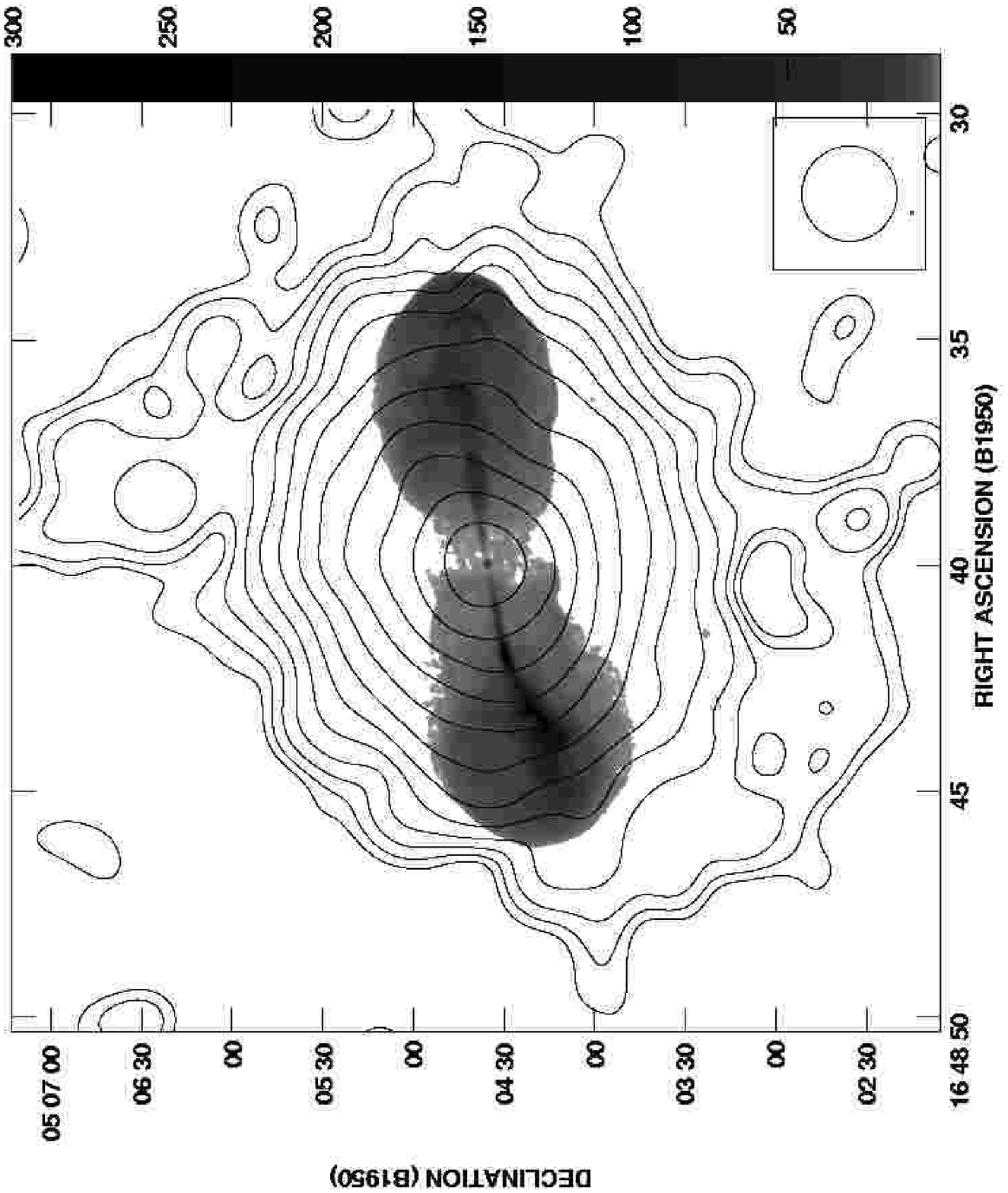}}
\end{picture}
\caption
{A contour map of the summed {\em ROSAT} PSPC and HRI 
image of the Hercules A cluster in the 0.5-2 keV band, overlaid on a
grey scale of the 20 cm radio map. The radio image is displayed with
a logarithmic transfer function, between 0.3 and 300 mJy. 
Contours are separated by a
ratio of $\sqrt{2}$. The resolution of the X-ray contour map is 
32$''\!$, and of the radio image, 1$''\!$.4.  The first contour
is at $2.94\times 10^{-10}$ W\,m$^{-2}$\,sr$^{-1}$,  and contours
are separated by a ratio of $\sqrt{2}$. The zero level has been
subtracted. The boxed circle gives the smoothing beam of the HRI data,
essentially the resolution of the X-ray map.}
\label{xr} 
\end{figure*}

To get a best-possible image of the cluster on arcminute scales, we
made a weighted average of the PSPC and HRI images. The PSPC data were
smoothed with a 20 arcsec gaussian as before, while the HRI image was
smoothed with a 31.5 arcsec gaussian to give approximately the same
resolution, allowing for the $\approx 25$ arcsec FWHM intrinsic PSF of
the PSPC. Both images were converted to instrument-independent brightness 
units. Weighting was by the inverse $\sqrt{N/2}$ variance in the
smoothed images, except that we did not let $N$ fall below the average
background value.  The larger smoothing of the HRI data makes the
latter about twice as sensitive in the cluster core, while at larger
radii the lower background of the PSPC means that the latter
dominates.  Fig.~\protect{\ref{xr}} shows contours of this weighted
image superposed on the same radio image as in Fig.~\ref{hrihigh},
this time displayed with a logarithmic transfer function to emphasise
the faint emission projected across the cluster core.

The total radio emission extent
is $\simeq$ 540 kpc, comparable to the size of the X-ray gas visible in
the images, but the radial profile in Fig.~\ref{pshri} shows that faint 
X-ray emission extends out to at least 2.2 Mpc radius.

\section{Model Fitting of the Surface Brightness Profile}
\label{surf}

Allowing for their different resolutions, the HRI and PSPC profiles
are in good agreement.
The profiles differ within $\sim 20$ arcsec
of the centre due to the lower resolution of the PSPC, and
are limited at large radii by the uncertainties in background 
level discussed in Sections~\protect{\ref{hi}} \& ~\protect{\ref{hri_data}}.

We tried fitting the surface brightness profiles with various
models, to find the electron density distribution in the
Hercules A cluster. Among these were a King or $\beta$-model 
\citep{King1962,Cavaliere78}, 
a de Vaucouleurs $r^{1/4}$ model and a combination
(sum) of both.  We confidently excluded the flat-topped `King-law'
profile, as it failed to describe the brightness spike
in the center of the cluster.
The de Vaucouleurs
model, as well as the model resulting from the sum of the latter with
the King model, also failed, this time at larger radii. A somewhat
better fit was obtained with an analytic `cooling flow' model 
\citep{Bertschinger.etal1986}, but this still failed to represent the
central peak. We therefore concluded that this peak was a discrete feature
and not just part of a continuous brightness distribution. 

We obtained
a good fit from a combination of a $\beta$-model with a central
point source.  We emphasise that, although the $\beta$-model was originally 
proposed as an approximation to the isothermal sphere, it provides an
accurate empirical fit to the emissivity in many `cooling flow' clusters
\citep{White.etal1997}. For the point source
we used the theoretical HRI PSF \citep{David.etal1998}, as
none of the background sources are bright enough to give a measure of the 
PSF width.  This does not give a perfect fit to the centre of the radial 
profile, not surprisingly, in view of the obvious elongation of the peak
in the image. 
Following Leahy \& Gizani (2001), we also tried fitting with a `blurred' PSF
to allow for aspect errors, 
but the improvement in the fit was only just significant at
the 5 per cent level, when fitting the HRI data alone; for the
better-constrained joint fit to HRI and PSPC, the improvement
was insignificant. Hence we use the nominal PSF in
the following.  The construction of the HRI PSF followed the method
of Leahy \& Gizani (2001), which allows for the finite pixel size of the
original image. For the PSPC, where accuracy is less critical because
the compact component is only weakly detected, we simply
evaluated the PSF model of \citet*{Hasinger.etal1995} at the centre of each
2.5-arcsec radial bin (assuming a photon energy of 1 keV). The results
are given in Table~\ref{kingfit}. Since the PSPC background model has a 
slight radial dependence, in Table~\ref{kingfit} we quote the value at the
field centre. $S$ is the point source flux and $I_0$ the central brightness
of the $\beta$-model.

\begin{table*}
\caption{The final results from the model fit to the
surface brightness profiles}
\begin{minipage}{\linewidth}
\def\footnoterule{\kern-3pt
\hrule width 2truein height 0pt\kern3pt}
\begin{center}
 
\begin{tabular}{lcccccc} \hline 
 
& S & $I_{0}$ & $r_{c}$ & $\beta$ & $\chi^{2}$ & Background \\ 
& ${\rm ct \, s^{-1}}$ &
${\rm ct\, s^{-1}\,sr^{-1}}$ & kpc & & & ${\rm ct\, s^{-1}\,sr^{-1}}$\\ \hline 
HRI only & $0.0095 \pm 0.0018$ & $9.6 \pm 0.9 \times 10^{5}$  & $117 \pm
19$ & $0.71 \pm 0.08$ & 25.5 (25 df) & $1.05 \times 10^5$\\
PSPC only & $0.0058 \pm 0.0048$ & $10.7 \pm 1.6 \times 10^{5}$  & $112 \pm 15$ 
& $0.72 \pm 0.04$ & 39.5 (27 df) & $9.59 \times 10^3$ \\ 
Combined fit & $0.0092 \pm 0.0014$ & $9.6 \pm 0.6 \times 10^{5}$ &
$121 \pm 10$& $0.74 \pm 0.03$  & 65.8 (56 df) & $9.60 \times 10^3$\\ 
\hline
 
\end{tabular}
\end{center}
\end{minipage}
HRI fluxes have been converted to the PSPC scale by multiplying by 2.86.
\label{kingfit}
\end{table*}

\begin{figure*}
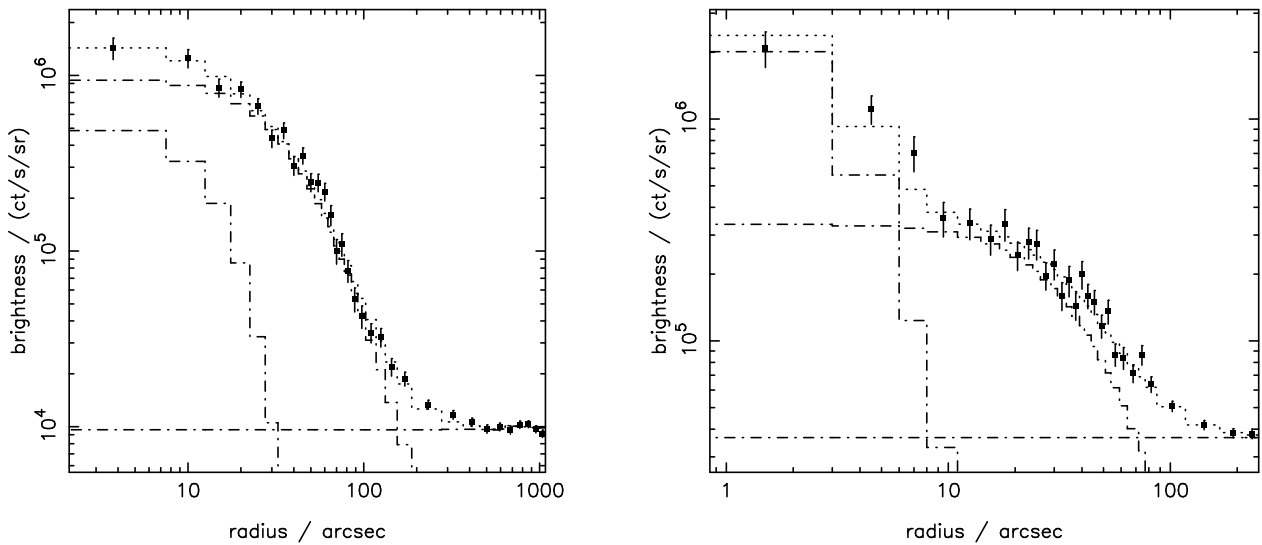

\centering
\setlength{\unitlength}{1cm}
 
\begin{picture}(17.5,7.5)
\put(-1.5,8.7){\includegraphics{ngizani6a.eps}}
\put(7.5,8.7){\includegraphics{ngizani6b.eps}}
\end{picture} 
\caption{The best fit compact source + $\beta$ model fit to the
radial surface-brightness profiles of the Hercules A cluster. In each
plot the combined model is given by the dashed line and the three
components of the model (PSF, $\beta$ model, and background) are
given by dash-dot lines.
(a) PSPC data. (b) HRI data.}
\label{kingplot} 
\end{figure*}

The PSPC profile fit was broadly consistent with that for the HRI, 
although the central peak is not clearly detected because 
the contrast against the cluster emission is much less with the lower
resolution. 
To get a best estimate of the cluster structure, we fitted both profiles
simultaneously, allowing the PSPC background to vary and keeping the
HRI background fixed. 
The combined best fit is given in Table~\ref{kingfit}. We use the
parameters found from this in our further analysis. The best fit
is shown plotted against the data from each instrument in Fig.~\ref{kingplot}.

The combined fit is excellent for the HRI data. The HRI data
contribute 25.6 to $\chi^2$ of the combined fit, with 29 bins. The
fit to the PSPC data is less good, $\chi^2 = 40.2$ with 32 bins. Within
a radius of 100 arcsec, the HRI data is as sensitive as the PSPC, and
the good fit to the former suggests that the poor PSPC fit is just a
random error. The PSPC does reveal excess emission compared to the
$\beta$ model at 200--400 arcsec. 

The value found for $\beta$ is within the range 0.5 to 0.8 
found for most clusters (\citealt{Jones.etal1984};
\citealt*{Cirimele.etal1997}). \citet{Siebert.99} find $\beta =0.63$ and a
core radius of 90 kpc (for our $H_0$), but this is based on fitting the
HRI data only, and only in a sector away from the radio lobes. We replicate
their result in Section~\ref{interaction}. The PSPC gives a much stronger
constraint on $\beta$, and at radii much larger than the radio 
lobes and likely unaffected by them, so it makes sense to average all
azimuthal angles.

To estimate the possible variability of the central source, we fitted
the HRI and PSPC data separately, fixing the parameters of the $\beta$
model and backgrounds to those found in the joint fit. The result is
not well constrained because the core is so weakly-detected by the
PSPC; we find an insignificant increase of $13 \pm 36$ per cent
between the PSPC and HRI observations.  As expected from the small
contribution of the core to the total flux, any power-law spectral
component was undetected in the {\em ASCA} and {\em BeppoSAX} 
spectra (Siebert et
al., Trussoni et al.); as Trussoni et al. point out, this does confirm
that the core is not heavily obscured. 

The integrated flux in the PSPC profile within 500 arcsec is $0.207
\pm 0.005$ ct s$^{-1}$, which slightly exceeds the combined flux in
the model, due to the deviations from the model noted above. To get our
best estimate for the total flux of the cluster, we subtract the
central source, apply a 3 per cent correction to infinite radius
(found analytically from our $\beta$ model fit), and an 11 per cent
correction to the total band (found by integrating the observed
spectrum), PI channels 11-223, giving $0.227 \pm 0.006$ ct s$^{-1}$,
corresponding to a 0.1--2.4 keV (absorbed) flux density of $3.04 \times
10^{-15}$ W m$^{-2}$, a (rest frame, unabsorbed) luminosity of
$3.00\times 10^{37}$~W, and a bolometric luminosity of $4.84 \times
10^{37}$ W (all calculated for our two-temperature spectral model).  
These figures are consistent with (but slightly lower
than) the values derived from the much shorter RASS observation by
Ebeling et al. (1998); they are in excellent agreement with the
results from the long {\em Einstein Observatory} exposure analysed by
White et al. (1997) (allowing for the different $H_0, q_0$ assumed in
these studies).\nocite{Ebeling.etal1998,White.etal1997} 

We saw above that relatively cool gas contributes 
$f_{\rm cool} \approx 15$ per cent of the 
cluster emission. This must have a density $\gta 4$ times higher than
adjacent hot gas for pressure balance, and is also likely a more efficient
emitter (i.e. higher $f(T,Z,z)$).
Thus {\em if} it is radially segregated (as in cooling-flow clusters), 
{\em then} it must nearly fill the cluster core; otherwise
we would expect a two-scale core in emission
(the central source cannot be the cooling-flow core, as its flux is less than
half that of the cool component). In fact 11--21 per cent of the cluster
emission comes from within 0.65--0.89 core radii for $\beta = 0.74$. 
Thus complete radial segregation is barely consistent, even if 
$f_{\rm cool}$ is at the upper end of the range (which we recall implies 
$\langle kT\rangle$ closer to 1 than 0.5 keV).

We have found the electron density from the emission 
measure, $EM$, and the conversion factor derived in
Section~\ref{pspcbaspe} (recall this assumes $kT = 4.25$ keV).  
The electron density $n_{e}$ of the $\beta$ model
cluster can be written as
\begin{equation}
n_{e} = 
\frac{n_{0}}{\left(1 + (r / r_c)^{2}\right)^{3\beta/2}}
\label{c2}
\end{equation}
Taking $n_e = 1.2n_H$, we find a central electron density $n_{0} =
1.0\times 10^{4}$ m$^{-3}$, suggesting a quite dense cluster, as
typical cluster central number densities are of
order $10^{3}$ m$^{-3}$ \citep{Jones.etal1984}. 
If we have a cooling flow, the central density is probably overestimated
by 30--50 per cent, as noted earlier; but the densities derived from our
model (with $n_0$ as quoted) should be correct at $r > r_c$, since the
model correctly predicts the brightness at all radii, and the cool
component makes at most a small contribution at large $r$.

\section{Discussion}
\subsection{Interaction of the radio lobes with the cluster gas}
\label{interaction}

We have looked for distortions in the cluster gas near the radio lobes, which
might reveal interactions between the gas and the supposedly expanding 
lobes. As a first step we made
radial profiles from the PSPC and the HRI data in four azimuthal sectors
centered on the X-ray compact source.
Two sectors were taken with position angles on the radio lobes (`on-lobe'
profile), with a full width of 90$^{\circ}$ each, at PA 98$^{\circ}$ and
$-82^{\circ}$. The other two sectors had position angles off the lobes
(`off-lobe' profile), with the remaining position
angles. Tables~\ref{sectors_on} and \ref{sectors_off} show the results
of the best compact source plus $\beta$ model fit to the on-lobe
and off-lobe profiles.
The free parameters were $\beta$ and the core radius $r_{c}$;
the X-ray flux from the compact source, the central
brightness, and the background were kept fixed to the values found
from the combined fit in Table~\ref{kingfit}.

\begin{table}
\caption{The results from the 3-component model fit to the
on-lobe profiles}
\begin{minipage}{\linewidth}
\def\footnoterule{\kern-3pt
\hrule width 2truein height 0pt\kern3pt}
\begin{center}
 
\begin{tabular}{lccc} \hline 
 
& $r_{c}$ & $\beta$ & $\chi^{2}$  \\ 
&
 kpc & & \\ \hline 
HRI only & $
149 \pm 19$ & $ 0.85 \pm 0.11$ & 28.5 (18 df) \\
PSPC only & $148 \pm 11$& $ 0.84 \pm 0.05$ & 23.3 (21 df) \\ 
\hline
 
\end{tabular}
\end{center}
\end{minipage}
\label{sectors_on}
\end{table}

\begin{table}
\caption{The results from the 3-component model fit to the
off-lobe profiles}
\begin{minipage}{\linewidth}
\def\footnoterule{\kern-3pt
\hrule width 2truein height 0pt\kern3pt}
\begin{center}
 
\begin{tabular}{lccc} \hline 
 
& $r_{c}$ & $\beta$ & $\chi^{2}$ \\ 
&  kpc & & \\ \hline 
HRI only & 
$98 \pm 13$ & $0.65 \pm 0.06$ &13.9  (13 df) \\
PSPC only &  $ 126 \pm 11 $ 
& $ 0.83 \pm 0.06$ & 39 (18 df) \\ 
\hline
 
\end{tabular}
\end{center}
\end{minipage}
\label{sectors_off}
\end{table}

\begin{figure*}
\centering
\setlength{\unitlength}{1cm}
 
\begin{picture}(17.5,7.5)
\put(1.5,7.8){\includegraphics{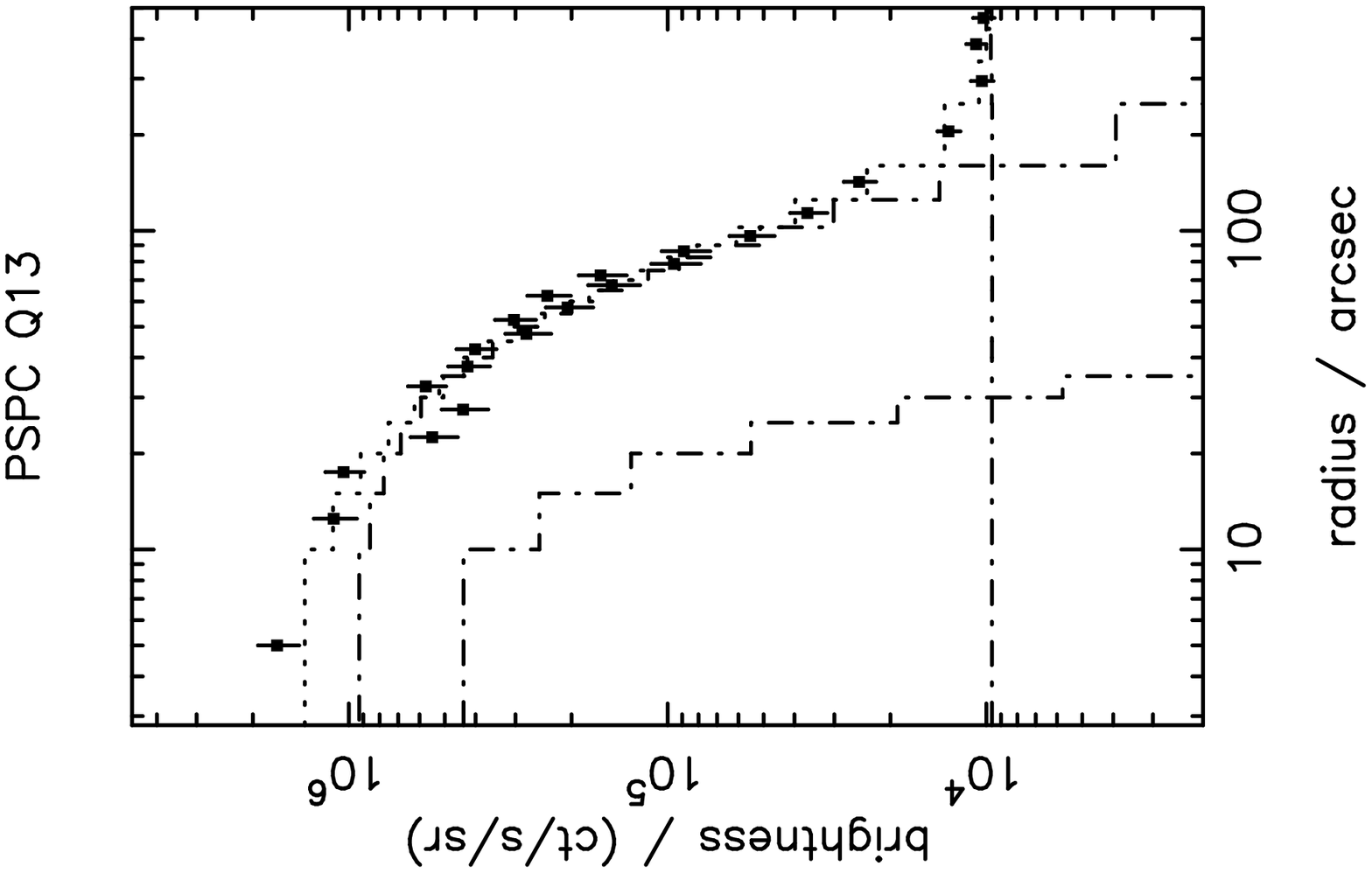}}
\put(9.,7.8){\includegraphics{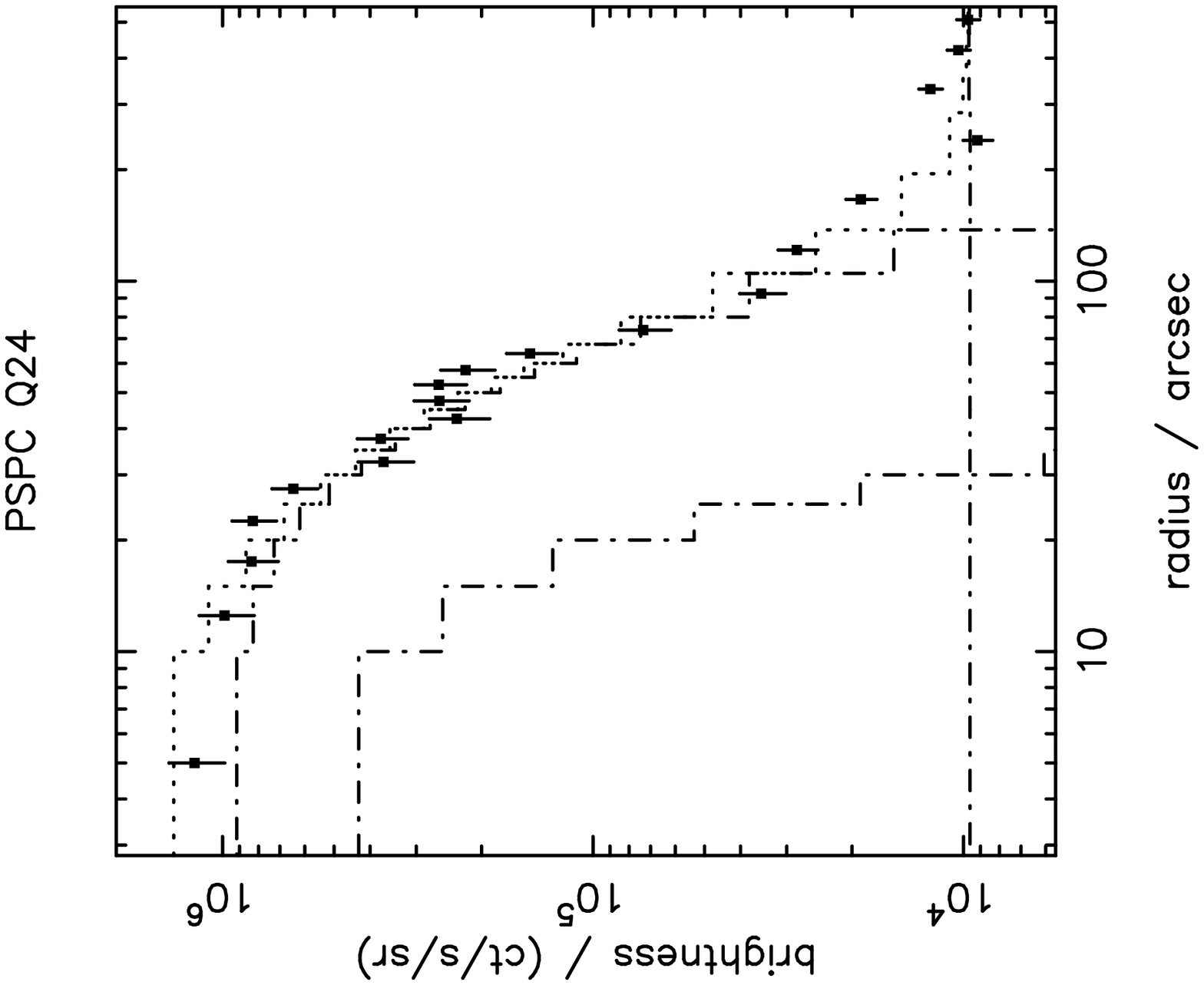}}
\end{picture} 
\caption{PSPC data: Radial profiles in sectors centered on- (left
panel) and off- (right panel) the radio lobes.}
\label{sectors1} 
\end{figure*}

\begin{figure*}
\centering
\setlength{\unitlength}{1cm}
 
\begin{picture}(17.5,7.5)
\put(0,7.8){\includegraphics{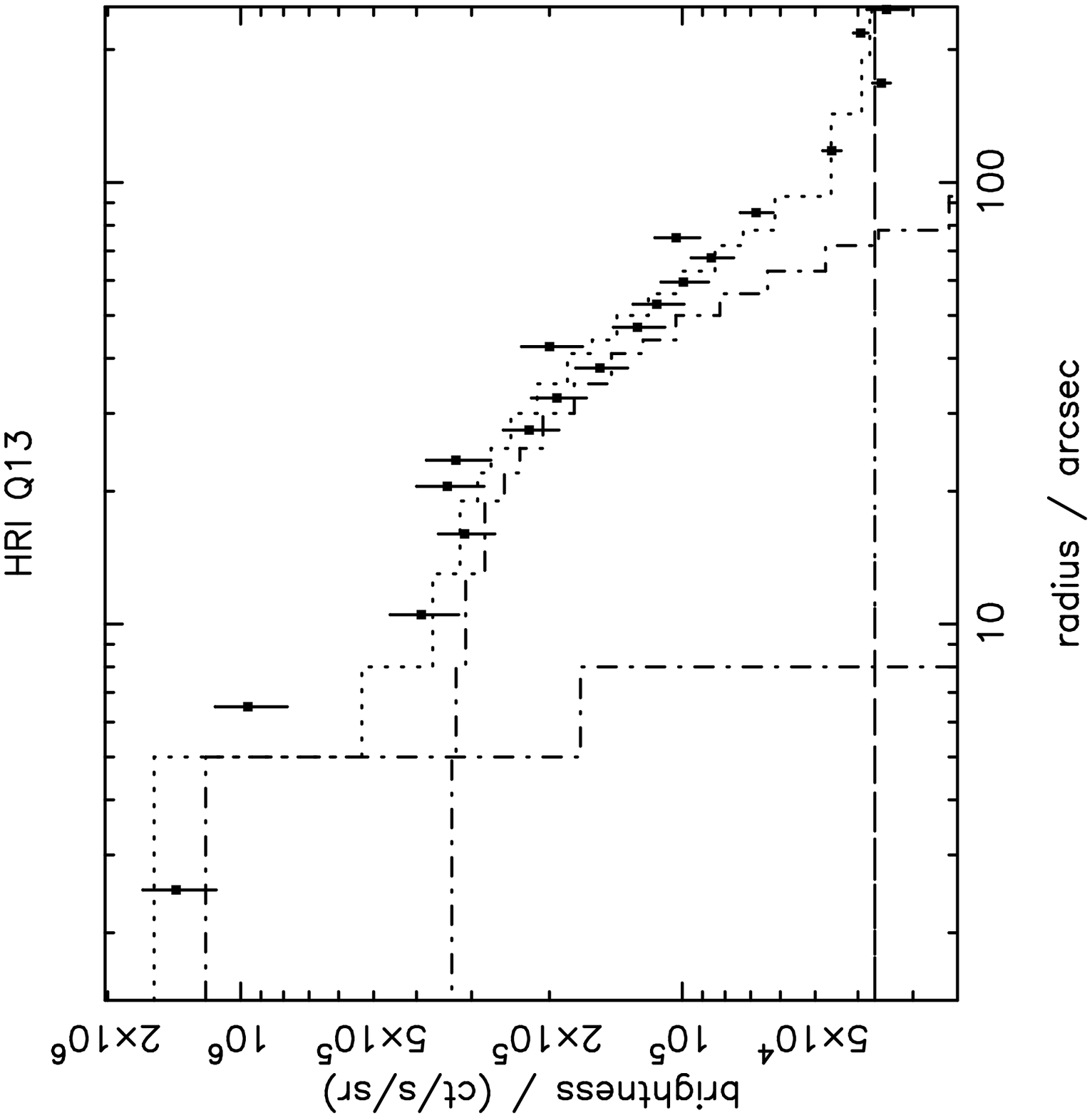}}
\put(8.5,7.8){\includegraphics{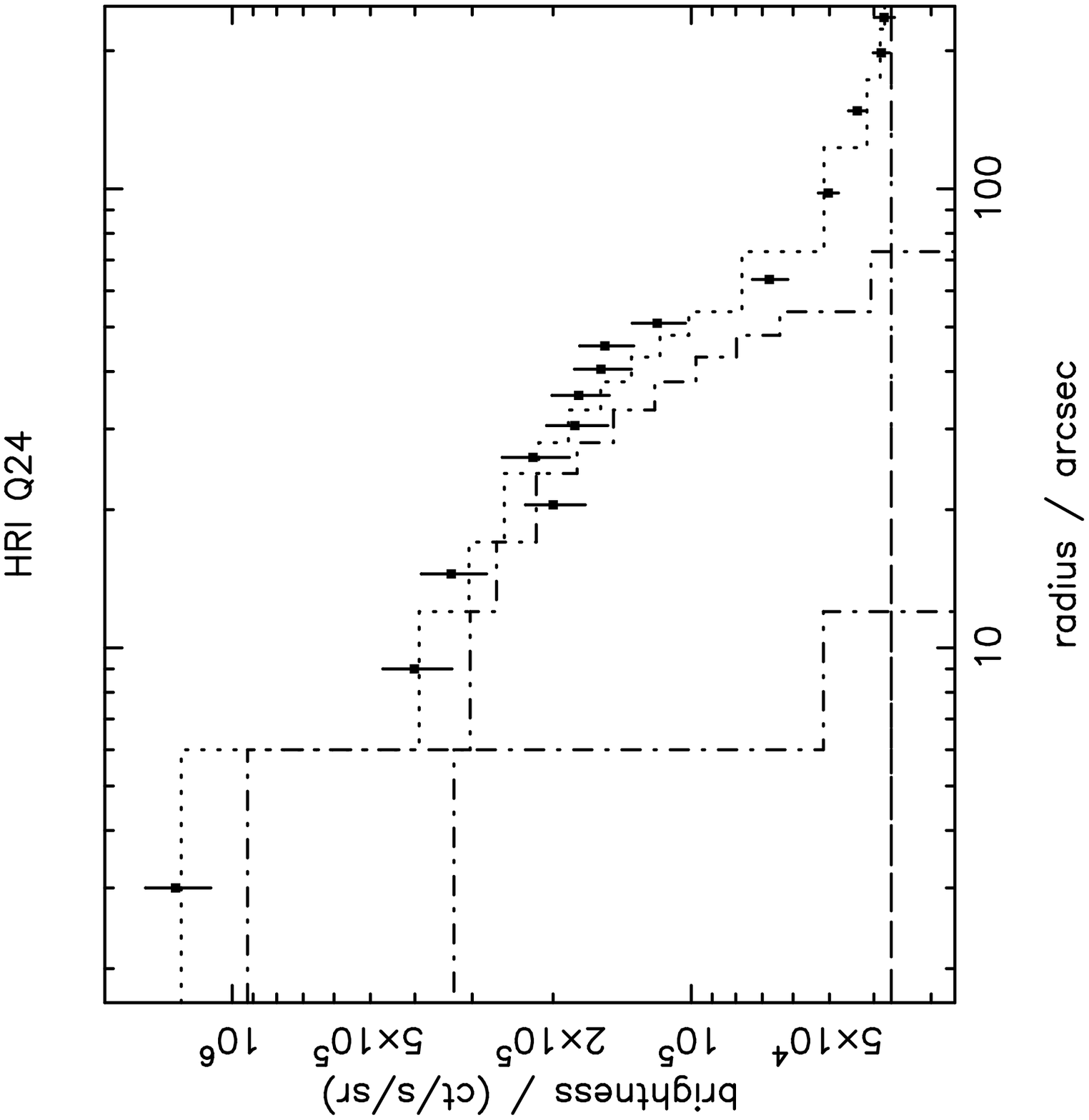}}
\end{picture} 
\caption{HRI data: Radial profiles in sectors centered on- 
(left panel) and off- (right panel) the radio lobes.}
\label{sectors2} 
\end{figure*}

Figs. \ref{sectors1} and \ref{sectors2} show the on-  and
off-lobe profiles for the PSPC and HRI respectively. In each
of the plots the combined model is given by dashed line and the three
components (PSF, $\beta$ model, and background) are given by dash-dot lines.

Table \ref{sectors_on},
from the profiles on the radio lobes, shows that PSPC and HRI
agree well, and show that the core of the gas distribution is
significantly elongated
in the direction of the lobes, as also visible in the images.
There is less good agreement transverse to the lobe axis 
(Table \ref{sectors_off}) suggesting that the profile
departs significantly from a $\beta$-model, especially at large
radii where only the PSPC is sensitive. This is partly due to the 
faint  excess emission to the north of the cluster visible in the PSPC
image (Fig.~\ref{xraypspc}). Given the better fit of the model to the HRI
data, we prefer the parameters derived from this fit for the inner 
regions of the cluster. If we take the off-lobe HRI fit as representing
the undisturbed cluster, we can interpret the on-lobe fit as suggesting
that the lobes have pushed some material out of the core, extending the
apparent core radius and steepening the profile at slightly larger radii.

We have seen that the X-ray emission is
elongated along the radio axis. The maximum-resolution HRI image
(Fig.~\ref{hrihigh}) shows discrete structures superimposed on the 
radio lobes, in particular arcs of emission perpendicular to the radio axis
near the ends of each lobe. The brighter, western arc lies just beyond
the brightest radio shell (or arc) embedded in the lobe. 
These features have been cited as evidence for a `radio/X-ray'
interaction by Siebert et al. (1999).\nocite{Siebert.99}
The expected brightness from the smooth model of Table~\ref{sectors_on}
at radii 74 and 88 arcsec (corresponding to the east and west X-ray arcs) is
$2.2\pm 1.5$ and $1.33 \pm 1.3$ HRI ct beam$^{-1}$, 
whereas the observed peaks are 5.7 and 6.8 ct beam$^{-1}$. 
(Because Poisson noise depends on the {\em
expected} counts, errors are properly attached to the model rather than to
the data). At low count levels, positive deviations are more probable
than for a Gaussian noise distribution, so the eastern `arc' is not very
significant, but the western arc seems to be real. The other structures 
noted by Siebert et al. are only marginally significant, 
except for the most interesting, the ridge extending west from the core;
unfortunately the latter is close enough to the core
that it may be an artefact of pointing errors. As the discussion by
Siebert et al. makes clear, 
there is no clear pattern in the relation between radio and X-ray
features; e.g. the western jet is roughly superimposed on the X-ray ridge
while the brighter eastern jet lies partly in an X-ray valley. Although
Seibert et al. read the structure in terms of shells of material around the
outer edges of the radio lobes, the enhanced emission is almost all seen
projected {\em inside} the outer boundary of the radio lobes, which
makes interpretation problematic. 
If these features are analogous to the enhancements around radio lobes
seen in some other objects (e.g. Vir~A, \citealt{Belsole01}; Hyd~A,
\citealt{Nulsen02}), then we would
expect them to be cooler than the ambient ICM. Indeed, they are plausible
candidates for the spectrally-detected $\approx 0.7$~keV phase.

\begin{figure*}
\centering
\setlength{\unitlength}{1cm}
 
\begin{picture}(15.5,15.5)
\put(0,0){\includegraphics{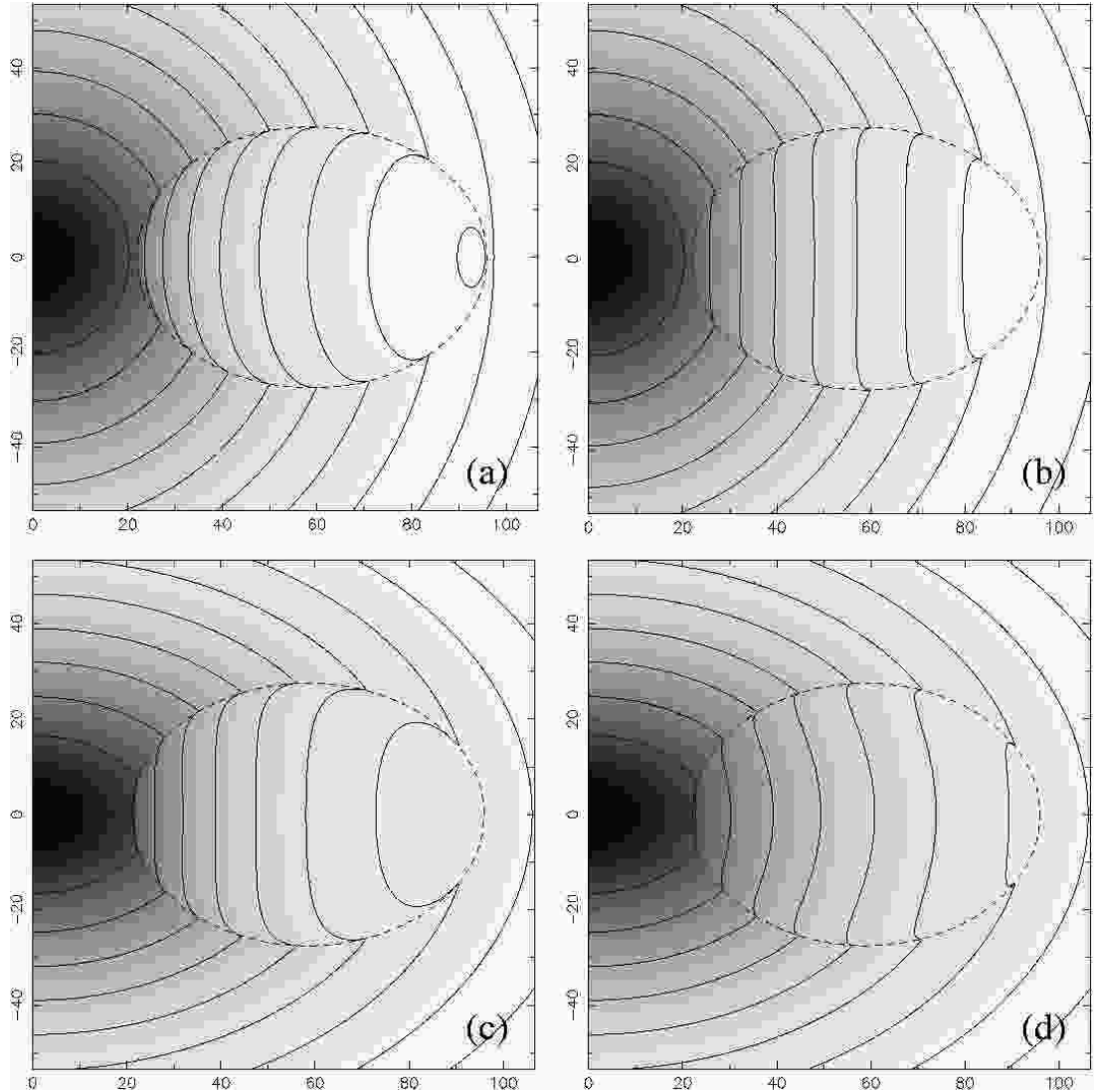}}
\end{picture}
\caption{Models of the X-ray emission from $\beta$-model clusters with
an ellipsoidal hole of zero emissivity, representing the radio lobe.
In all cases the hole has major and minor projected axes 74 and 55 arcsec,
its end is projected 96 arcsec from the cluster centre: the dashed 
ellipse shows its outline. The greyscale is linearly proportional to the
model brightness, and the solid lines are X-ray isophotes separated
by ratios of $\sqrt 2$. We
take $\beta = 0.74$ for all the cluster models. (a) Spherical
cluster with $r_c = 43$ arcsec; radio axis inclination $i=90^\circ$.
(b) as in (a), but $i=50^\circ$. (c) Elliptical cluster model with projected
core radii 53 and 35 arcsec. Cluster and radio axis inclination both 
$50^\circ$. (d) as in (c) but cluster assumed to be in the plane of the
sky, while the radio axis remains at $i=50^\circ$.}
\label{hole}
\end{figure*}

Notably lacking in the X-ray images are any clear depressions in the
cluster gas associated with the radio lobes, of the kind found
in Cyg A \citep{Carilli.94,Wilson.etal00}, 
Per A \citep{Bohringer.et93,Allen.et00}, and several other DRAGNs
\citep{McNamara.00,FJ2001,YWM2002}.
Given that the
lobes of Her A are comparable in size to the cluster core, this is 
surprising, at least on the standard interpretation that the lobes are
relativistic `bubbles' which displace the thermal gas. Certainly, we
cannot interpret the radio bridges as components of a cylindrically 
symmetric bubble passing straight through the cluster centre; this puzzle
is discussed in detail in Paper II.  Here we focus on the `bulbs', that
is, on the bright, roughly elliptical outer lobes. Fig.~\ref{hole} shows
the predicted X-ray brightness from our $\beta$-models for the Her~A cluster,
but with an ellipsoidal hole of zero X-ray emission matched in size to
the radio bulbs. In Fig.~\ref{hole}a, the DRAGN
is assumed to be in the plane of the sky. 
The brightness is reduced by a factor of almost two at the lobe
centre. 
In Fig.~\ref{hole}b, we assume the DRAGN is inclined at
$50^\circ$ to the line of sight, as estimated in Paper III from our
depolarization data. 
In Fig.~\ref{hole}c we take the cluster as elliptical, with the projected 
major and minor core radii as given by our sector fits. We assume the
cluster is aligned with the radio lobes in three dimensions, i.e. both
have $50^\circ$ inclination. 
Finally, in Fig.~\ref{hole}d the cluster elongation is in the plane 
of the sky, but the radio source remains
at $i = 50^\circ$. This gives very small distortions of the X-ray isophotes,
although this geometry is contrived, implying as it does that the 
observed alignment between cluster and DRAGN is merely a chance 
projection effect.

The low signal-to-noise ratio in our HRI data, the low
resolution of the PSPC, and the presence of the enhancements 
discussed above (which will partially mask any depressions), make it
hard to exclude any of these models, although the case shown in
Fig.~\ref{hole}a is certainly disfavoured. 
Our HRI image (Fig.~\ref{hrihigh}) shows a ridge of emission 
along the northern edge of the western lobe, reminiscent of the generic
features in Fig.~\ref{hole} (but actually {\em more} prominent than in any
of our models); but any ridges along the other lobe edges must be much
weaker, at the level of the random fluctuations in the image.
In short, our data are consistent with models in which the radio `bulbs'
are indeed empty of thermal gas, especially if we are correct in inferring 
that the radio axis is substantially out of the sky plane.
Higher resolutions and deeper observations with {\em Chandra} should 
clarify the details of this interaction (Wilson et al., 2003, in prep.).

\subsection{Confinement of the radio lobes}
\label{confinement}

As the radio lobes of Hercules A lack anything resembling 
hotspots, it seems likely that they are expanding at around the sound
speed,
and so are confined by the thermal pressure of the ambient ICM, rather
than by strong shocks.  Therefore a measurement of the thermal pressure 
should be a measurement of the energy density
of the lobes. The lobes are sharply bounded (e.g. Fig.~\ref{xr}), suggesting
little entrainment, so the lobe energy should be dominated by 
relativistic particles and fields. We therefore have the opportunity to
test the minimum energy estimate $u_{\rm min}$ with the more direct 
measurement from the external pressure.

As in Section~\ref{interaction}, we focus on the bulbs rather than the faint
inner lobes. Although the ends of the
bulbs are at almost exactly the same distance from the centre, the
more elliptical western bulb reaches closer to the middle and so on
average is closer to the centre.
Following the methodology of Leahy \& Gizani
(2001) \nocite{Leahy.etal1999} we have calculated the minimum 
pressure ($P_{\rm min} = u_{\rm min}/3$) and the thermal pressure at the
deprojected distance of the eastern and western bulbs of Hercules
A, assuming $i \approx 50^\circ$. 
For the calculation of the thermal pressure, $P_{\rm th} = 1.9 n_e kT$,
we have used the results of the combined fit of the surface brightness
profile model (see Table~\ref{kingfit}).
The results are listed in Table~\ref{pressures1}: 
$R$ is the deprojected distance of the bulb centres, 
while $B_{me}$ is the minimum energy magnetic field. In addition to the
conventional minimum energy parameters we also list the values $B_{MS}$
and $P_{MS}$ found using the methodology presented in 
Myers \& Spangler (1985),\nocite{Myers.etal1985} 
that is, assuming the electron energy spectrum extends down to
a Lorentz factor of 10.

\begin{table}
\caption{Minimum energy parameters and the thermal pressure}
\begin{minipage}{\linewidth}
\def\footnoterule{\kern-3pt
\hrule width 2truein height 0pt\kern3pt}
\begin{center}
 
\begin{tabular}{lcccccc} \hline
Component & R & $B_{me}$ & $P_{\rm min}$ & $P_{\rm th}$ & 
$B_{MS}$ & $P_{MS}$ \\  
         & arcsec &  nT   & pPa  & pPa &  nT & pPa \\ \hline
East bulb &  84 &  0.79 & 0.19 & 2.3 & 1.99  & 1.05 \\ 
West bulb &  71 &  0.73 & 0.17 & 3.0 & 1.86  & 0.92 \\
\hline
 
\end{tabular}
\end{center}
\end{minipage}
\label{pressures1}
\end{table}

As we can see from Table~\ref{pressures1},
the radio lobes of Hercules A would be significantly
underpressured with the standard minimum energy assumptions for the
radio plasma: electron/positron composition (i.e. no protons), 
filling factor unity and a low-energy cutoff corresponding to 10 MHz emission.
Thus minimum energy is a severe underestimate of the actual 
energy content of the lobes, by an order of magnitude. 

Such a deficit in the minimum energy has been established for a long time for
FR\,I DRAGNs \citep*[e.g.][]{Feretti1992}, but FR\,I jets and tails were
often modelled as turbulent flows with substantial entrainment, so the 
result could be attributed to mixture with the 
ambient gas and/or  to a low filling factor for the radio plasma. 
This was ruled out in some FR\,Is by the the discovery of X-ray cavities
\citep{Bohringer.et93,McNamara.00}, and by deficits in cases like Her~A,
where the lobes are sharply bounded \citep{Hardcastle.2000,LG2002}.  
This points to an error in the microphysical assumptions, 
as discussed in detail by \citet{Leahy.etal1999} for 3C\,388.
In that case we argued that the 
extra energy is mostly in the form of particles, and as a result the 
magnetic field should be below equipartition and therefore does not play an 
important role in the lobe dynamics.  The results of the Myers \& Spangler
calculations show that an alternative option is available for Her A:
equipartition can be (almost) preserved if the particle energy spectrum 
extends down to trans-relativistic energies. The difference arises 
essentially because the steeper spectrum of Her A gives a larger increase 
in energy on extrapolating to low Lorentz factors.

\subsection{Cooling and heating of the cluster gas}

The intracluster gas is densest in the core of the cluster and
therefore the radiative cooling time $t_{\rm cool}$, due to the emission
of the observed X-rays is shortest there \citep*{Fabian.etal1991,Fabian1994}.
For isobarically cooling gas
\begin{eqnarray}
t_{\rm cool} &=& {\frac{5}{2} kT (n_H + n_e + n_{He}) \over 
\Lambda(T,Z) n_e n_H}
= {5.75\, kT \over \Lambda(T,Z)\, n_e}  \\ 
&\approx& 10^{11} \left(\frac{n_e}{10^{3} \, {\rm m^{-3}}}\right)^{-1}\,
\left(\frac{T}{10^{8}\,{\rm  K}}\right)^{\frac{1}{2}} {\rm yr} \nonumber
\end{eqnarray} 
where $\Lambda(T,Z)$ is the cooling function (see Leahy \& Gizani
2001).  The last form is the approximation of \citet{Sarazin1986}, which
begins to become inaccurate for temperatures below 3 keV.  For the
centre of the Hercules A cluster the cooling time is 6.4 Gyr
if the hot phase dominates, but  $\approx 2$ Gyr if the
core is dominated by the cool phase. Since this is
less than the age of the universe, this would normally imply the
presence of a cooling flow \citep{Sarazin1986}. The cooling radius,
i.e. the radius at which the cooling time equals $10^{10}$ yr, is
$\approx 90$~kpc, somewhat less than the core radius, consistent with
the relatively good fit to the $\beta$ model. 

\citet{Peres.etal98} find that the core radii of cooling flows
clusters is generally smaller than in non-cooling flow clusters, with
the division around 130 kpc, which also suggests that Her A has a 
cooling flow. However, the Peres et al. core radii are fitted values for
the (assumed isothermal) cluster {\em mass} distribution, and are not 
directly comparable to {\em gas} core radii derived from $\beta$-model fits.

It now appears that in many or all clusters, some agency prevents fully 
formed cooling flows from developing, so that little gas cools to much 
less than half the ambient cluster temperature (see 
\citealt{MP2001,Peterson.etal2003} for recent analyses). 
Energy input from DRAGNs is a popular candidate for this agency, 
\citep[e.g.][and references therein]{Churazov2002}. Attention has mostly
been focussed on weaker, quasi-continuous AGN activity rather than
major episodic outbursts as in Her~A, but the latter may also play an important
r\^{o}le \citep[e.g.][]{David2001}.

Although our results show that the normal state of the Hercules A 
cluster is a cooling flow, a region a few times the size of the core is now
affected by the expansion of the radio lobes, and 
it is easy to see that the `waste' energy
supplied to the environment by the DRAGN
currently far outweighs the energy loss from X-ray radiation.
If we assume quasi-isobaric (and so subsonic) expansion, 
so that the cluster pressure distribution
is not disturbed, the average rate of work done by the DRAGN is
$dW/dt = \int_{\rm lobes} p\,dV/\tau$ where $\tau$ is the lifetime. As in
Section~\ref{interaction} we focus on the outer `bulbs' of the lobes,
and we model these as prolate ellipsoids inclined at $i \approx 50^\circ$;
numerical integration gives a `pressure content' of
$\sim 5 \times 10^{53}$~J.
Writing $\tau = \tau_8 \times 10^8$~yr, this gives
$dW/dt \sim 1.6 \tau_8^{-1} \times 10^{38}$ W. 
The combined kinetic power of the jets $K = 4\, dW/dt$, as it
equals the enthalpy flux into the lobes (we assume that the
lobes are dominated by relativistic plasma, so $u = 3P$).  
We can also 
parameterize the problem in terms of the radiative efficiency of the 
DRAGN. With the radio luminosity $L_{\rm radio} =  3.8 \times 10^{37}$~W,
\[ \varepsilon = L_{\rm radio} / K \approx 0.06 \tau_8, \]
and we can write $dW/dt = L_{\rm radio} / 4\varepsilon $.
Estimates for the lifetimes of powerful DRAGNs are typically in the range
$0.1 \lta \tau_8 \lta 1$, corresponding to an efficiency of 
0.6--6 per cent for Her A. 
Given that the radio luminosity of the DRAGN and the X-ray luminosity of
the entire cluster are almost the same, the cluster gas is currently gaining
energy at 3--30 times its radiative loss rate.  
\citet*{Reynolds.etal2001} point out that initially the work done may appear
as potential energy (some of the gas is lifted out of the cluster core);
in the long term much of this will be re-cycled to heat via
turbulent dissipation.  In supersonic expansion, of course, there will
be at least some direct heating by shocks. 
In either case, any cooling of the cluster will be temporarily reversed.
When the radio outburst is over, the cluster will return to its cooling state. 
At present, most of the mechanical energy is being deposited well outside
the cluster core, and it is possible that the core is already
cooling again, even while the outer part of the cluster is heated. 
 
If the efficiency is
around 1 per cent, and the AGN is periodically active with a duty cycle 
of around 4 per cent, then on average radiative cooling of the cluster is
balanced by mechanical heating.
The cooling time of the cluster would then provide a rough measure of the 
time since the last outburst. Such a scenario appears to suffer from 
fine tuning \citep{FMNP2001}, but, as suggested by \citet{Churazov2002},
feedback via accretion onto the AGN from the cooling gas could naturally 
regulate the cycle of heating and cooling.

Although it is most likely that Her A is expanding transonically, 
it is interesting to consider
how compression of the cluster gas behind the bow shock of a DRAGN 
will affect the cooling time.  We have found how the Mach number changes the
cooling time, by using the \citet{Sarazin1986} approximation,
and the change in temperature and density as a function of shock Mach number
\citep{Landau.etal1959}.
For shock normal Mach numbers ${\cal M} < 6.5$
the cooling time is initially somewhat reduced, but only by up to around 
40 per cent, which is not nearly enough to make radiative losses important 
on the timescale of the AGN ($\lta 10^8$~yr). Over the longer 
term, the material will re-expand adiabatically to a condition with
a longer cooling time than before. Higher Mach numbers will increase the
cooling time from the start, but for ${\cal M} \gta 10$ the shocked
gas will be too hot to be visible in X-rays. We note that in the case
of Hercules A the pre-shock central cooling time would have been
$< 10^{10}$ yr whatever the Mach number, i.e. the apparent
cooling flow cannot be an artefact of the DRAGN. 

\subsection{The magnetic field in the cluster core}

\begin{figure}
\centering
\setlength{\unitlength}{1cm}
 
\begin{picture}(8.5,8.5)
\put(-0.6,-2){\includegraphics{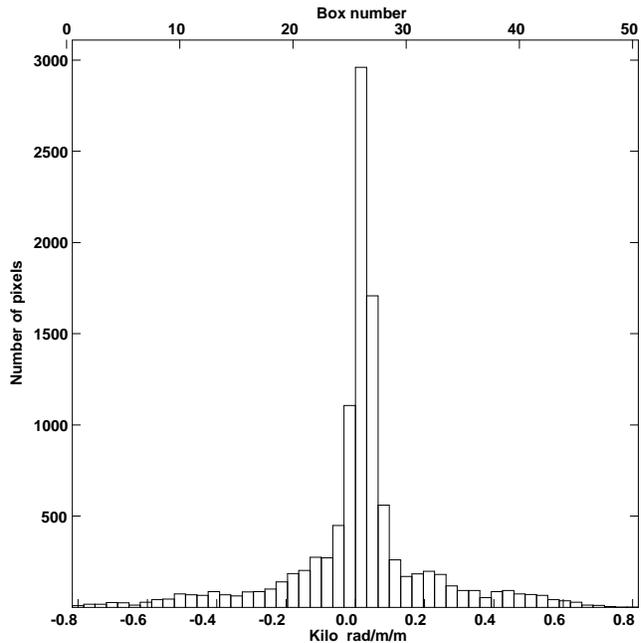}}
\end{picture} 
\caption{Histogram of the rotation measure for all significant pixels within
43.4 arcsec of the radio core of Her A, from our radio data (Paper III).}
\label{srm} 
\end{figure}

In Paper III we estimate the strength of the intracluster
magnetic field as a function of radius, 
using our X-ray data and the Faraday dispersion from our
radio data. In this section we attempt a cruder estimate,
using the dispersion of the rotation measure $\sigma_{RM}$ in the
central $r_{c}$ = 43.4 arcsec. We follow the analysis by 
\citet{Garrington.etal1991},
who show that the expected $\sigma_{RM}$ for a source behind
a Faraday screen with a tangled
magnetic field with rms $B$ nT (independent of radius), is given by
\begin{equation}
\sigma_{RM} = 0.0081 \left[ \int (n_e B_\|)^2 l \, dz \right]^{1/2}
\end{equation}
where $z$ is along the line of sight, $l$ is the coherence length,
(both in kpc), $n_e$ is electron density in m$^{-3}$, and $B_\|$ is
the component of the field parallel to $z$, equal to $B/\sqrt{3}$ for
isotropic tangling.  For a source at the center of a cluster whose radial
profile is described by a $\beta$-model this becomes
\citep[e.g.][]{Felten.96}
\begin{equation}
\sigma_{RM}  = \frac{0.0081 B_{\|} n_0 r_{c}^{1/2}l^{1/2}}
                    {(1 + r^2/r_{c}^2)^{(6\beta-1)/4}}
\sqrt\frac{\pi^{1/2} \Gamma(3\beta -0.5)}{2 \Gamma(3\beta)},
\label{sigrm}
\end{equation}
where $n_0$ is the central electron density and $\Gamma$ is the Gamma
function.  Figure~\ref{srm} is a histogram of the rotation measure for
all significant pixels in Her A within the core radius, from our radio
data (Paper III). We find $\sigma_{RM} = 214$ rad m$^{-2}$, and the
coherence length $l$ near the centre is about 5 arcsec or 14 kpc
(Paper III). We take a fiducial value of $r = 80$~kpc (28.6 arcsec),
since most of the pixels with valid rotation measures are towards the
outer part of the region.  Then, using $n_0$, $\beta$ and $r_c$ from
our combined HRI and PSPC fit, the estimated $B \simeq$ 0.2 nT.
Hence the magnetic field in the central region of the cluster
of Hercules A is lower than in extreme cooling flow clusters (e.g. the
3C\,295 cluster, \citealt{Allen.00}), but within the typical range for
clusters in general \citep{CT2002}.

\subsection{The central source}

As we have seen in Section~\ref{surf} the improvement of fitting a
blurred PSF to the central peak is insignificant. This means that both
a nominal PSF and a blurred PSF are both good fits of the weak central
peak in the HRI image (for example see Fig.~\ref{hrihigh}). 
All we can say is that we have a central source with a count rate
of 0.009 ct s$^{-1}$ in our hard band (Table~\ref{kingfit}). The peak
could be either an AGN or a galaxy-scale cooling flow. 
We will discuss each case separately:

\subsubsection{The central source as an AGN}

We have already noted that there is no sign of an obscured AGN in the
high-energy {\em BeppoSAX} spectrum \citep{Trussoni.00}, so we assume
a power-law model with only Galactic absorption.  Assuming a typical
photon index of $\Gamma = 1.9$, the unabsorbed 0.1--2.4 keV luminosity
of the central source is then $2.0 \times 10^{36}$~W.  At 5 GHz, the
radio core luminosity is $7.1 \times 10^{23}$~W~Hz$^{-1}$. 
The point source appears a
few times over-luminous in X-rays relative to the correlation between
$L_{X}$ and radio core power for galaxies 
\citep[corrected to our $H_0$]{Sieb96}, although it
lies within the general scatter.  We note that Siebert et al's
correlation is formed mostly by broad-line objects, while Her A is a
low excitation radio galaxy \citep{Tadhunter.etal1993,JR97}, with a
steep spectrum (optically thin) radio core at GHz frequencies (Paper
II).  Although X-ray variability is expected in an AGN we have not
detected any during the interval of the two and a half years between
our PSPC and HRI observations.  AGN X-ray emission could be
synchrotron or inverse Compton. \citet{Hardcastle.2000a} have argued
that inverse Compton is required, at least in FR\,Is, if the unified
model of low-power radio galaxies and BL Lacs is correct.

\subsubsection{The central source as a cooling flow}

The extra broadening by the PSF that fits the data best corresponds to the
nominal PSF convolved with a gaussian of FWHM 5.0 arcsec. Therefore
the width of the cooling flow peak, if any, would have to be around 5.0 
arcsec or less.  In contrast the cluster core radius found from the fits 
is 43~arcsec, an order of magnitude larger. This would give 
the cluster emission a `prussian hat' look: essentially the central
peak is on the scale of the central cD galaxy rather than the cluster. 
Small central peaks like this are sometimes seen in cooling flows. The 
X-ray luminosity is just greater than the largest
value for an individual galaxy in the {\em Einstein Observatory\/} sample
of \citet*{EFK95}, but this sample excluded cluster-centre galaxies. 
This suggests that if the central source is thermal emission it relates to
the cluster and/or the AGN, rather than solely to the host galaxy.  In this
case the central source, at around 5 per cent of the cluster luminosity,
would presumably contribute to the cool component detected in the spectrum, 
but could not produce all of it, as previously noted.

\section{Conclusions}

Our {\em ROSAT} PSPC and HRI X-ray observations of the intracluster gas in
the Hercules A cluster have revealed an extended X-ray emission, extending
to radius 2.2 Mpc (PSPC data) and a weak central peak (HRI data). 

Based on discrepancies between single-temperature fits to the PSPC,
{\em ASCA} and {\em BeppoSAX} spectra, 
we argue that the intracluster gas contains at 
least two phases, dominated by a $\approx 4$ keV component (seen by {\em ASCA}
and {\em BeppoSAX}), but with around 15 per cent of the emission in the 
{\em ROSAT} band from material at 0.5--1 keV.  This model significantly
improves the PSPC fit compared to one with a single temperature.
The 0.1--2.4 keV (absorbed) flux density is $3.0\times 10^{-15}$ W m$^{-2}$,
and the bolometric luminosity is $4.8 \times 10^{37}$ W. There is no 
evidence for absorption above the Galactic value.

We have detected X-ray emission coming from a compact source at
the centre of the cluster, with a 0.1--2.4 keV luminosity of $2 \times
10^{36}$~W. Our data does not distinguish between an AGN and galaxy-scale
thermal emission (perhaps a central cooling-flow spike).
The central electron density (excluding the central source) 
is $n_{0} \approx 1.0 \times 10^{4}$ m$^{-3}$.  The central
cooling time is in the range 2--6 Gyr, depending on whether or not the
cool gas phase is concentrated in the centre. The cooling radius is 90 kpc. 
The cluster core should therefore be a cooling flow if we ignore the
effect of the DRAGN. Indeed, most of the time, the DRAGN will be absent, 
as its life-time is certainly short compared to the age of the Universe.
But at present the DRAGN, which is several times larger than the cooling
radius, is almost certainly depositing more energy into this region of the 
ICM than is being lost radiatively.

Our best combined (PSPC and HRI) fit to the surface brightness
profile, with a model consisting of a PSF, a $\beta$ model and a
background, gives a $\beta$ parameter of $0.74 \pm 0.03$, typical
for clusters of galaxies. The radio lobes are largely positioned
beyond the X-ray core radius, allowing for projection, 
so they are expanding essentially into a power-law
atmosphere with density falling as $r^{-3\beta} \sim r^{-2.22}$, quite
close to the $r^{-2}$ profile needed to give the lobes a self-similar
structure \citep{Falle1991}. 

The thermal pressure at the deprojected distance of the radio lobes is
an order of magnitude larger than the minimum pressure of the
lobes. Thus the minimum energy in the lobes is a severe underestimate
of the actual energy content.  This seems to be typical for DRAGNs,
of both FR types, although only in a few cases can it be established as
clearly as in Her~A.

Two features of the X-ray emission may result from the interaction of the
cluster gas with the radio lobes. 
The region around the core of the cluster is clearly
elongated along the radio axis, as revealed by our on- and off-lobe
profiles. In addition there are a number of apparently discrete X-ray
enhancements projected on or around the radio lobes; however some of these
may be artefacts or due to poor photon statistics in our ROSAT data.
Unlike some other well-studied DRAGNs, Her~A as yet does not show
X-ray holes coincident with the radio lobes. 
But we have shown that the such holes could be too shallow to detect in
our data, due to projection effects:
the main lobes being displaced along the line of sight away from the
X-ray bright region around the cluster core. 
Deep observations with {\em Chandra} should reveal some sign of these holes, 
if present, and also clarify the nature of the `enhancements'.
 
\section*{Acknowledgments}

We thank Megan Donahue for helpful discussions.

Nectaria Gizani would like to acknowledge: PPARC for funding her fees
for three years to carry out her PhD work at Jodrell Bank at the
University of Manchester; The grant PRAXIS XXI/BPD/18860/98 from the
Funda\c c\~ao para a Ci\^encia e a Tecnologia, Portugal for the
post-doctoral fellowship; The current post-doctoral grant under
contract 332 from the State Scholarships Foundation (IKY), Greece. NG
is grateful to Jodrell Bank Observatory for the support of a
post-doctoral fellowship.
 
This research has made use of the NASA/IPAC Extragalactic Database
(NED) which is operated by the Jet Propulsion Laboratory, Caltech,
under contract with the national aeronautics and space administration.
It has also made use of NASA's Astrophysics Data System Abstract
Service (ADS).

We thank the UK ROSAT Data Archive Centre at the Department of Physics
and Astronomy, Leicester University and the ROSAT group. We used the
{\sc xspec} and {\sc asterix} software packages, and thank their
creators and maintainers.

\bibliography{3c348}

\end{document}